\documentclass[conference]{IEEEtran}

\ifCLASSINFOpdf
  \usepackage[pdftex]{graphicx}
\else
\fi

\usepackage{amsmath}
\usepackage{amssymb}
\usepackage{amsthm}

\usepackage{listings}
\usepackage[frozencache,cachedir=./minted-cache]{minted}
\usepackage{xcolor}

\usepackage{url}
\usepackage[hidelinks]{hyperref}

\newtheorem{theorem}{Theorem}
\newtheorem{proposition}[theorem]{Proposition}

\theoremstyle{definition}
\newtheorem{definition}{Definition}

\begin{document}

\title{Onion-Routed Multi-Circuit Key Establishment\\
       for Quantum-Resilient Sessions}

\author{
\IEEEauthorblockN{Tushin Mallick\IEEEauthorrefmark{1}}
\IEEEauthorblockA{Cisco Research}

\and
\IEEEauthorblockN{Ashish Kundu}
\IEEEauthorblockA{Cisco Research}

\and
\IEEEauthorblockN{Ramana Kompella}
\IEEEauthorblockA{Cisco Research}
}

\maketitle

\footnotetext[1]{This work was carried out when the author was at Cisco Research, San Jose, CA, during Summer Internship 2025. He is currently at Northeastern University, Boston, MA.}

\begin{abstract}
Public-key primitives that today anchor session-key establishment---RSA,
Diffie--Hellman, and elliptic-curve cryptography---reduce to integer
factorization or discrete logarithm and are therefore vulnerable to Shor's
algorithm on a sufficiently capable quantum computer.
The \emph{harvest-now, decrypt-later} (HNDL) threat model turns this future
capability into a present liability: ciphertext archived today can be
decrypted retrospectively once a cryptographically relevant quantum computer
becomes available. We propose a session-key establishment scheme that
distributes a freshly generated key as multiple, independently encrypted
fragments across distinct, ephemeral Tor circuits between an onion-service
proxy and an onion-service client. Reconstruction requires every fragment;
each fragment travels its own per-bundle circuit established via a
\texttt{NEWNYM} signal. The security argument rests on the standard
end-to-end correlation bound for onion routing: an adversary controlling a
fraction of Tor relays must independently deanonymize every fresh circuit to
correlate the fragments belonging to one session, and the per-fragment
probability of success decays multiplicatively in the number of fragments.
We implement the design as a Flask-based prototype on AWS EC2, with both the
proxy and the client deployed as Tor onion services, and measure end-to-end
key-establishment latency. The implemented prototype completes a key
establishment in 13--20\,s on average (7--50\,s including tails), of which
approximately 88\% is attributable to Tor-related delay---a cost we discuss
in the context of the privacy-versus-responsiveness trade-off.
\end{abstract}

\section{Introduction}
\label{sec:intro}


Modern secure communication on the open Internet rests on a small set of
public-key primitives---RSA, Diffie--Hellman (DH), and elliptic-curve
cryptography (ECC)---whose security reduces to the conjectured intractability
of integer factorization or discrete logarithms. Shor's algorithm solves both
problems in quantum polynomial time~\cite{shor1997}, so a sufficiently capable
quantum computer breaks all three. While no such machine yet exists,
contemporary risk analyses treat its eventual arrival as an engineering
problem rather than a theoretical one~\cite{mosca2018}.

The operational urgency comes from the \emph{harvest-now, decrypt-later}
(HNDL) threat model: an adversary intercepts and archives ciphertext today
and defers decryption until a cryptographically relevant quantum computer
(CRQC) becomes available. For data whose confidentiality must hold for years
or decades, present-day classical key establishment is therefore already
inadequate. Mosca's inequality $X + Y \geq Z$ captures the resulting risk,
where $X$ is the required confidentiality lifetime of the data, $Y$ is the
migration runway to quantum-resistant cryptography, and $Z$ is the time until
a CRQC is fielded~\cite{mosca2018}. The U.S. National Institute of Standards
and Technology (NIST) has finalized the first post-quantum cryptography (PQC)
standards---FIPS\,203 (ML-KEM), FIPS\,204 (ML-DSA), and FIPS\,205 (SLH-DSA)
in August 2024~\cite{fips203,fips204,fips205}---and major content delivery
networks and browsers have begun deploying hybrid classical/PQC handshakes
in production~\cite{westerbaan2023}.

\textbf{Motivation. }Two observations motivate the present work. First, even after PQC standards
are universally available, the wire-level migration is gated by a long tail
of constrained, embedded, and operational-technology endpoints that cannot
readily accept new cryptographic primitives~\cite{nistir8547}. During that
migration window, classical key establishment will continue to be performed
on links subject to HNDL collection. Second, while the cryptographic content
of a session-key exchange is one defensible surface, the \emph{metadata} of
the exchange is another. Knowing which two endpoints established a session,
and when, is independently sensitive in many settings; it also creates the
linkage an HNDL adversary needs in order to correlate a future decryption
with an identified pair of communicants. Defenses that protect only the
former leave the latter intact.

We address both observations by routing the key-establishment traffic
through the Tor anonymity network~\cite{dingledine2004} and by distributing
the established key as multiple independently encrypted fragments across
\emph{separate, ephemeral} Tor circuits. This approach is an instance of the general multi-path quantum resistance framework that is proposed and described elsewhere and is out of the scope of this paper (We will add a reference to that work when a publicly available reference is available).  The circuit-creation interface
exposed by the Tor controller (the \texttt{NEWNYM} signal accessible via the
\texttt{stem} library~\cite{stem}) makes it cheap to obtain a fresh path
through the network for each fragment bundle. Both the client and the
key-distribution proxy expose Tor onion services, so neither party reveals
its network location to the other or to any single relay on the path.\\

\textbf{Our Contributions. }This paper is the anonymous-paths companion to the Aquaman
protocol~\cite{aquaman2025}. We make the following contributions.
\begin{itemize}
  \item We instantiate the deferred
        anonymous-paths variant over the public Tor network: both
        proxy and client are exposed as Tor onion services, and a
        fresh circuit per fragment bundle is obtained via the
        \texttt{NEWNYM} control signal.
  \item We prove a multi-circuit linkability
        bound: an adversary surveilling a fraction $f$ of Tor relays
        correlates every one of the $n$ ephemeral circuits with
        probability at most $f^{2n}$ (Proposition~\ref{prop:decay}),
        and formalize the property as anonymity-preserving
        reconstruction (Definition~\ref{def:apr}).
  \item We implement the design on AWS EC2 and
        measure end-to-end key-establishment latency of 13--20\,s,
        of which roughly 88\% is attributable to Tor.
\end{itemize}

\textbf{Scope. } 
We do not propose a new key encapsulation
mechanism, and we do not claim that the onion-routing layer alone provides
post-quantum confidentiality of the fragment payload: per-fragment encryption
is performed under the recipient's classical public key, and an HNDL adversary
that captures and later decrypts a single fragment recovers that fragment's
plaintext. The contribution is on \emph{linkability}: distributing the
fragments across independent, anonymous Tor circuits raises the cost of
identifying which decrypted fragments must be combined to reconstruct a key
and, hence, of attributing the reconstructed key to a specific pair of
endpoints. The scheme composes with a post-quantum KEM on the wire as a
defense-in-depth measure; we discuss this composition but do not implement
it in the current prototype.
The proposed approach in this paper is based on the general multi-path quantum resistance framework that is proposed and described elsewhere and is out of the scope of this paper (We will add a reference to that work when a publicly available reference is available).

\textbf{Outline. }Section~\ref{sec:background} reviews the relevant cryptographic and
networking background. Section~\ref{sec:related} surveys related work in
post-quantum migration, multi-path key establishment, and anonymous
communication. Section~\ref{sec:threat} formalizes the threat model.
Section~\ref{sec:system} describes the system design.
Section~\ref{sec:eval} reports the prototype implementation and empirical
evaluation. Section~\ref{sec:discussion} discusses limitations and the
usability--security trade-off with conlcusion in Section~\ref{sec:conclusion}.

\section{Background}
\label{sec:background}

This section reviews the technical material on which the design rests:
the quantum threat to asymmetric cryptography, the Tor anonymity network
and its onion-service protocol, and the use of secret sharing for key
distribution. We deliberately omit introductory cryptographic material
that is well covered elsewhere and focus on the points that bear directly
on the design choices in Section~\ref{sec:system}.

\subsection{The Quantum Threat to Asymmetric Cryptography}
\label{sec:bg:quantum}

Shor's algorithm provides a quantum polynomial-time solution to integer
factorization and to the discrete-logarithm problem in finite fields and on
elliptic curves~\cite{shor1997}. Because RSA, classical DH, and ECC reduce
to instances of these problems, a CRQC of sufficient scale breaks all
three. Symmetric primitives are comparatively less affected: Grover's
algorithm yields only a quadratic speed-up against generic key
search~\cite{nistir8105}, so doubling key sizes (e.g., AES-256) is widely
regarded as adequate against quantum adversaries. The brunt of the
quantum threat therefore falls on the \emph{key-establishment} layer of
secure protocols, which is the layer this work targets.

The HNDL threat operationalizes this future capability today. Ciphertext
collected from any classically-keyed channel can be retained indefinitely
and decrypted in retrospect once a CRQC is fielded. Mosca's inequality
$X+Y\geq Z$ states the resulting exposure: organizations whose data
shelf-life $X$ plus migration runway $Y$ exceeds the time $Z$ to CRQC are
already exposed regardless of when $Z$ actually arrives.

NIST's PQC effort has produced four standards as of this writing: ML-KEM
(FIPS\,203), ML-DSA (FIPS\,204), SLH-DSA (FIPS\,205), and a forthcoming
Falcon-based signature standard, with HQC selected in March 2025 as a
fifth, code-based backup KEM~\cite{fips203,fips204,fips205,hqc2025}.
NIST IR\,8547 sets a deprecation horizon of 2030 for quantum-vulnerable
algorithms at security strengths $\leq 112$ bits, and full disallowance
by 2035~\cite{nistir8547}. Hybrid constructions that combine a classical
KEM (typically X25519) with a PQ KEM (typically ML-KEM-768) are the
operational consensus for the transition window~\cite{westerbaan2023}.

Three frictions persist. \emph{First}, the deployed PQC primitives are
younger than the classical primitives they replace and have already
exhibited unexpected weaknesses: SIKE, a fourth-round NIST candidate,
was broken in classical polynomial time by Castryck and Decru in
2022~\cite{castryck2023}, and side-channel and fault-injection attacks
have recovered keys from Kyber/ML-KEM and Dilithium
implementations~\cite{ravi2024}. \emph{Second}, PQC ciphertexts and
signatures are substantially larger than their classical counterparts and
interact poorly with bandwidth-limited or fragmentation-sensitive
protocols~\cite{nistir8547}. \emph{Third}, the migration window between
today and the 2030/2035 horizon is precisely the window the HNDL adversary
exploits, regardless of how rapidly endpoints upgrade.

\subsection{Tor and the Onion-Service Protocol}
\label{sec:bg:tor}

Tor is a volunteer-operated, low-latency anonymity network in which a
client routes its traffic through a sequence of relays such that no single
relay learns both the source and the destination~\cite{dingledine2004}.
Onion routing layers encryption: each relay decrypts one layer and
forwards the remainder. A standard client circuit consists of three
relays---a guard, a middle relay, and an exit---and the client extends the
circuit one hop at a time using the ntor handshake, a one-way authenticated
key agreement based on ECDH over Curve25519~\cite{goldberg2013ntor,
torspec216}. After ntor, each hop derives per-direction symmetric keys via
HKDF-SHA256; the resulting onion-encrypted cells are protected with
AES-128 in counter mode~\cite{degabriele2018onionae}.

\textbf{Onion services. }Onion services (formerly ``hidden services'') allow a server to accept
connections without revealing its IP address~\cite{torproject_hs,
torspec_rend}. The server publishes a signed descriptor to a distributed
hash table maintained by the directory authorities and announces a small
set of \emph{introduction points} reachable via three-hop circuits. A
client retrieves the descriptor, builds its own three-hop circuit to a
relay it selects as \emph{rendezvous point}, sends an \texttt{INTRODUCE1}
cell through an introduction point, and the service then completes a
circuit to the rendezvous point, which splices the two halves into one
bidirectional stream~\cite{torspec_rend, skerritt2023}. The complete
client-to-service path therefore traverses six relays: three picked by the
client, including the rendezvous point, and three picked by the
service~\cite{torproject_hs}.

Two properties of onion services matter for our design.
\begin{enumerate}
  \item \emph{Bidirectional anonymity.} Neither party learns the other's
        IP address; addressing is done by \texttt{.onion} hostnames derived
        from the long-term Ed25519 identity key of the service.
  \item \emph{Self-authenticating endpoints.} The \texttt{.onion} address
        is the encoded master identity key (plus a version byte and
        checksum), so successful connection to an address authenticates the
        service end-to-end without external certificate
        authorities~\cite{torspec_rend}.
\end{enumerate}

\textbf{Controller interface and ephemeral circuits}
Tor exposes a control protocol over a local socket configured by the
\texttt{ControlPort} directive. Applications interact with this control
plane programmatically through the \texttt{stem} Python library or any
other implementation of the control specification~\cite{stem}. Among the
control signals, \texttt{NEWNYM} instructs Tor to use a fresh circuit for
subsequent stream attachments---paths to a given destination are randomly
re-selected from the relays advertised in the current
consensus~\cite{stem_faq}. The Tor Project explicitly cautions that
\texttt{NEWNYM} does not guarantee a new exit IP address or a globally
disjoint path, and that aggressive circuit churn imposes load on the
network~\cite{stem_faq}; we revisit this caveat in
Section~\ref{sec:discussion}.

\textbf{Threat surface. }Tor does not defend against an adversary that controls or surveils both
ends of a circuit simultaneously. The standard threat is the
\emph{end-to-end correlation attack}: by matching packet timing and volume
patterns at the entry and the exit of a circuit, an adversary observing
both can identify the circuit's pair of endpoints with high
confidence~\cite{murdoch2007, nasr2018deepcorr, surveycorrelation}. The
classical Tor analysis~\cite{dingledine2004} models the per-circuit
correlation probability under independent relay sampling as approximately
$(a/P)^2$, where $a$ is the number of compromised relays out of $P$
total. We use this bound in the security argument of
Section~\ref{sec:threat}.

\subsection{Secret Sharing and Multi-Path Distribution}
\label{sec:bg:ss}

The classical primitive for splitting a secret into independently useless
shares is Shamir's $(t,n)$-threshold secret sharing
scheme~\cite{shamir1979}: any $t$ shares reconstruct the secret, while
fewer than $t$ shares reveal nothing in an information-theoretic sense.
Our design uses an $(n,n)$-style split---all fragments are required for
reconstruction---rather than a true polynomial-interpolation Shamir
scheme. We make this choice deliberately for the following reasons.
\begin{itemize}
  \item Partial-availability tolerance is unnecessary in the design: every
        fragment is dispatched once over a per-bundle Tor circuit and the
        client either receives all fragments or fails the session and
        retries. Threshold tolerance would add bandwidth without changing
        the security argument.
  \item The adversarial model already provides the unavailability
        guarantee we need: a fragment that does not arrive at the client
        is not a confidentiality failure but a liveness failure.
\end{itemize}
The choice nonetheless places our work in the broader lineage of multi-path
key distribution. Barnett and Phoenix~\cite{barnett2011} treat intermediate
relays in a QKD relay network as trusted eavesdroppers and apply secret
sharing across multiple physical paths through the relay graph; a recent
IETF draft develops multi-path secret sharing for QKD key relay over IP
networks~\cite{li2026qkdmpss}. Costea \emph{et al.}~\cite{costea2018ompk}
explore multipath Diffie--Hellman variants and demonstrate that naive
share-then-DH constructions are vulnerable when adversaries can pre-agree
shares on multiple paths. Our setting differs from both: shares are not
the public key material itself but ciphertext fragments of an already
generated symmetric key, dispatched over independent Tor circuits whose
anonymity properties---not whose physical-medium diversity---are the source
of adversarial cost.

\section{Related Work}
\label{sec:related}

Five lines of prior work are most relevant to a Tor-routed, multi-circuit
key-establishment scheme: hybrid classical/PQC handshakes, post-quantum
migration of Tor itself, multi-path key establishment over heterogeneous
or threshold-shared paths, anonymity systems and end-to-end correlation
attacks, and edge-side post-quantum key exchange. We describe each in
turn.

\subsection{Hybrid Classical/PQC Key Establishment}
\label{sec:rel:hybrid}

The mainstream operational response to cryptanalytic uncertainty around
new PQC primitives is the hybrid construction: combine a classical KEM
with a PQ KEM so that the derived shared secret is secure if either
component holds. KEMTLS, introduced by Schwabe, Stebila, and
Wiggers~\cite{schwabe2020kemtls}, replaces signature-based authentication
in TLS\,1.3 with KEMs and is structurally amenable to such hybridization.
The Open Quantum Safe project~\cite{stebila2017oqs} packages liboqs and
prototype integrations into mainstream cryptographic stacks; it has been
the substrate for many of the experimental hybrid TLS deployments in the
literature. Production deployments have followed: Cloudflare reports that
over a third of human-initiated HTTPS traffic on its network already uses
hybrid post-quantum key agreement, with X25519+ML-KEM-768 the standard
suite~\cite{westerbaan2023}. Specifications for the same hybrid group
have been ratified by the IETF and integrated by JDK, Chrome, and other
client stacks~\cite{jdk527}. Hybrid handshakes provide an algorithmic
hedge against the failure of a single cryptographic primitive while
leaving the surrounding protocol, endpoint, and channel structure
unchanged.

\subsection{Post-Quantum Migration of Tor}
\label{sec:rel:tor-pqc}

Several lines of work have studied replacing the ntor circuit-extension
handshake with a post-quantum or hybrid construction. Ghosh and
Kate~\cite{ghosh2015postquantumtor} introduce HybridOR, a one-way
authenticated key exchange that is secure under either the gap
Diffie--Hellman assumption or the ring-LWE assumption and is compatible
with Tor's existing public-key infrastructure. Subsequent
work~\cite{schanck2015tor} formalizes the security of hybrid
circuit-extension handshakes and benchmarks them on
constrained-device hardware. The Tor Project has issued specification
proposals \#249 and \#263 for cell-size widening, both subsequently
superseded by proposals \#269 and \#340 covering hybrid KEM integration
and cell packing respectively~\cite{torspec269,berger2025tor}.
Berger, Lemoudden, and Buchanan~\cite{berger2025tor}
survey the cryptographic primitives at each layer of the Tor stack and
quantify the performance overhead of replacing ntor with a hybrid
KEM-based construction (e.g., ML-KEM combined with X25519). The line
concerns the cryptography \emph{of the circuit-extension handshake
itself} rather than application-layer composition of Tor circuits.

\subsection{Multi-Path Key Establishment and Multi-Secret Sharing}
\label{sec:rel:multipath}

Barnett and Phoenix~\cite{barnett2011} described how to secure a
quantum-key-distribution relay network by treating intermediate relays as
trusted eavesdroppers and applying secret sharing across multiple
physical paths through the relay graph; an adversary recovering the
shared key must compromise every relay on every path. The QKD-networking
literature has continued this thread, developing multi-path key forwarding
across QKD networks with partial trust in relays and recent
zero-trust spatiotemporal diversification schemes for end-to-end
QKD~\cite{li2026qkdmpss,qkdnetperspective}.

In the classical setting, Costea \emph{et al.}~\cite{costea2018ompk}
analyze opportunistic multipath key exchange and show that naive
share-then-Diffie--Hellman variants are vulnerable when the adversary can
coordinate replies across paths; their construction motivates careful
design of which quantity is split and which is signed. Multipath TCP and
analogous transport-layer protocols~\cite{costea2018ompk} distribute
application data across multiple network paths; as a side effect, an
on-path observer of a single subflow recovers only a fragment of the
data. None of these constructions targets metadata privacy: the paths
are typically disjoint IP routes through the same underlying Internet,
and observers at the endpoints see the full fan-out structure of the
exchange.

The recent Aquaman proposal~\cite{aquaman2025} is a transparent-proxy
architecture supporting four operating modes---PQC offloaded to the
proxy, classical multi-path fragmentation across heterogeneous physical
media with an optional private proxy-pool variant, QKD-based key
delivery via SKIP or ETSI GS~QKD~014, and classical/PQC hybrid
handshakes---of which the first two are implemented. The multi-path
mode distributes ciphertext fragments of a session key, each encrypted
under the recipient's classical public key, across heterogeneous media
(Wi-Fi, Bluetooth, NFC, cellular, Ethernet), and Aquaman formalizes the
maximum recovery probability as decaying $(B/d)^{n}$ in the diversity
dimension $d$, with $B$ the adversary's per-type compromise budget and
$n$ the fragment count. A complementary private proxy-pool variant
routes the request through a pool of trusted proxies and bounds
single-session entry/exit correlation as $(a/P)^{2}$, matching the
standard onion-routing
analysis~\cite{dingledine2004,murdoch2007}; the detailed
anonymous-paths protocol is deferred to a separate work.

\subsection{Anonymity Systems and Correlation Attacks}
\label{sec:rel:anon}

Chaum's mix network~\cite{chaum1981mix} introduced the layered-encryption
relay construction that underpins all modern onion-routing systems.
Tor~\cite{dingledine2004} adapted this idea to interactive, low-latency
traffic using telescoping circuits, perfect forward secrecy, directory
authorities, and configurable exit policies. The dominant practical
threat against Tor is the end-to-end correlation attack, in which an
adversary observing both the entry and the exit of a circuit matches
timing and volume patterns to identify the
endpoints~\cite{murdoch2007,nasr2018deepcorr,surveycorrelation,
attacksontor}. The classical Tor analysis~\cite{dingledine2004} models
the probability that a circuit's entry and exit relays both fall under an
adversary controlling a fraction $f = a/P$ of relays as approximately
$f^2$ under independent sampling, and recent work~\cite{rector2025}
demonstrates that learning-based correlators continue to extend the
practical reach of this attack. The Loopix mix network~\cite{loopix2017}
addresses the timing channel directly with Poisson mixing and cover
traffic, accepting bandwidth and small added latency as the price of
resistance to a global passive adversary.

Onion services double the cryptographic anonymity but also double the
latency: the rendezvous protocol requires the client and the service to
each maintain three-hop circuits to a common rendezvous
point~\cite{darkhorse2023, torproject_hs}. The latency of onion-service
key establishment, rather than its cryptographic content, has been the
focus of several proposed performance
optimizations~\cite{darkhorse2023,wilms2008}.

\subsection{Post-Quantum Cryptography at the Network Edge}
\label{sec:rel:edge}

A growing body of work places post-quantum cryptographic capability at
the network edge---in CDN points-of-presence, cloud gateways, or
trusted-network boundary devices---rather than at every endpoint.
Industrial deployments of edge PQC TLS by Cloudflare, Google, AWS, and
others use hybrid X25519+ML-KEM ciphersuites~\cite{westerbaan2023}.
Amiriara, Mirmohseni, and Tafazolli~\cite{amiriara2025} formalize an
edge-computing PQC framework for resource-constrained IoT devices in
which devices offload heavyweight cryptographic operations to a
post-quantum edge server and use physical-layer techniques to protect the
device-to-edge link. Aquaman~\cite{aquaman2025} adopts the same edge
posture explicitly: its transparent proxy is deployed at an edge router
or gateway of a trusted network and intercepts client traffic at the
network layer, so that quantum-resistant capability is added to a
network without modifying the client devices it serves.
\section{Threat Model and Assumptions}
\label{sec:threat}

We formalize the adversary against which the design must hold and the
trust placed in each system component. The adversary's capabilities are
drawn from the standard onion-routing
literature~\cite{dingledine2004,murdoch2007} augmented with future
quantum cryptanalytic capability; the resulting model is the one against
which the security analysis of Section~\ref{sec:security} works.

\subsection{Adversary Model}
\label{sec:threat:adversary}

We consider a polynomial-time adversary $\mathcal{A}$ with the following
capabilities.

\textbf{Network observation. }$\mathcal{A}$ passively observes traffic at one or more vantage points
in the Internet, capturing ciphertext and packet-level metadata. This
subsumes the HNDL capability: archived ciphertext is retained
indefinitely and later inspected once $\mathcal{A}$ acquires the
cryptanalytic ability to decrypt it.

\textbf{Relay surveillance. }$\mathcal{A}$ may surveil or operate a subset $\mathcal{S} \subseteq
\mathcal{R}$ of the Tor relay set, with $|\mathcal{S}| = a$ out of
$|\mathcal{R}| = P$ relays in the consensus. On a relay in $\mathcal{S}$
the adversary observes inter-relay traffic (including under future
quantum capability that defeats the classical hop-to-hop encryption).
$\mathcal{A}$'s surveillance set is fixed in advance of any given
request and is not adaptively reconfigured during a single key
establishment~\cite{dingledine2004,murdoch2007}.

\textbf{Quantum cryptanalysis. }$\mathcal{A}$ is endowed, either at the time of observation or at some
future decryption time, with sufficient quantum computational capability
to break any classical asymmetric primitive whose security reduces to
integer factorization or to the discrete-logarithm problem (RSA, DH,
ECC). For symmetric primitives we assume only Grover-bounded attacks;
AES-256 and SHA-2/3 at $\geq 256$ bits remain quantum-secure under
standard assumptions~\cite{nistir8105}.

\subsection{Trust Assumptions}
\label{sec:threat:trust}

The QKMS is trusted with respect to the confidentiality of the
generated session key, since it samples the key in plaintext at
generation time; its trust footing is therefore comparable to that of a
certificate authority or a key-management appliance in a contemporary
enterprise PKI. The onion-service proxy is trusted to forward fragments
faithfully but \emph{not} to view plaintext key material: per-fragment
encryption is performed under the recipient's public key before the
bundle leaves the QKMS, so an honest-but-curious proxy that records
every bundle that traverses it never sees the session key in plaintext.
The Tor network itself is trusted only in the standard onion-routing
sense: at least one relay on each circuit is honest, and the adversary
surveils no more than a fraction $f < 1$ of relays.

\subsection{Threat}
\label{sec:threat:nongoals}

  The following threats are outside the scope of this paper and the approach described herein, and we assume these threats are addressed by means that are in the literature or in practice.
\begin{itemize}
  \item Compromise of the QKMS or of the endpoint's private key.
  \item An adversary that simultaneously deanonymizes \emph{every}
        fresh Tor circuit used during a single key establishment.
  \item Active denial-of-service attacks that prevent the client from
        receiving all $n$ fragments. These reduce \emph{availability}
        but not confidentiality.
  \item Side-channel attacks on endpoint hardware (timing, power, EM).
  \item Confirmation attacks in which the adversary has out-of-band
        reason to believe that two specific endpoints are communicating
        and merely seeks high-confidence confirmation. Tor itself does
        not defend against this class~\cite{torproject_hs}.
\end{itemize}
\section{System Design}
\label{sec:system}

\begin{figure*}[t]
  \centering
  \includegraphics[width=0.85\linewidth]{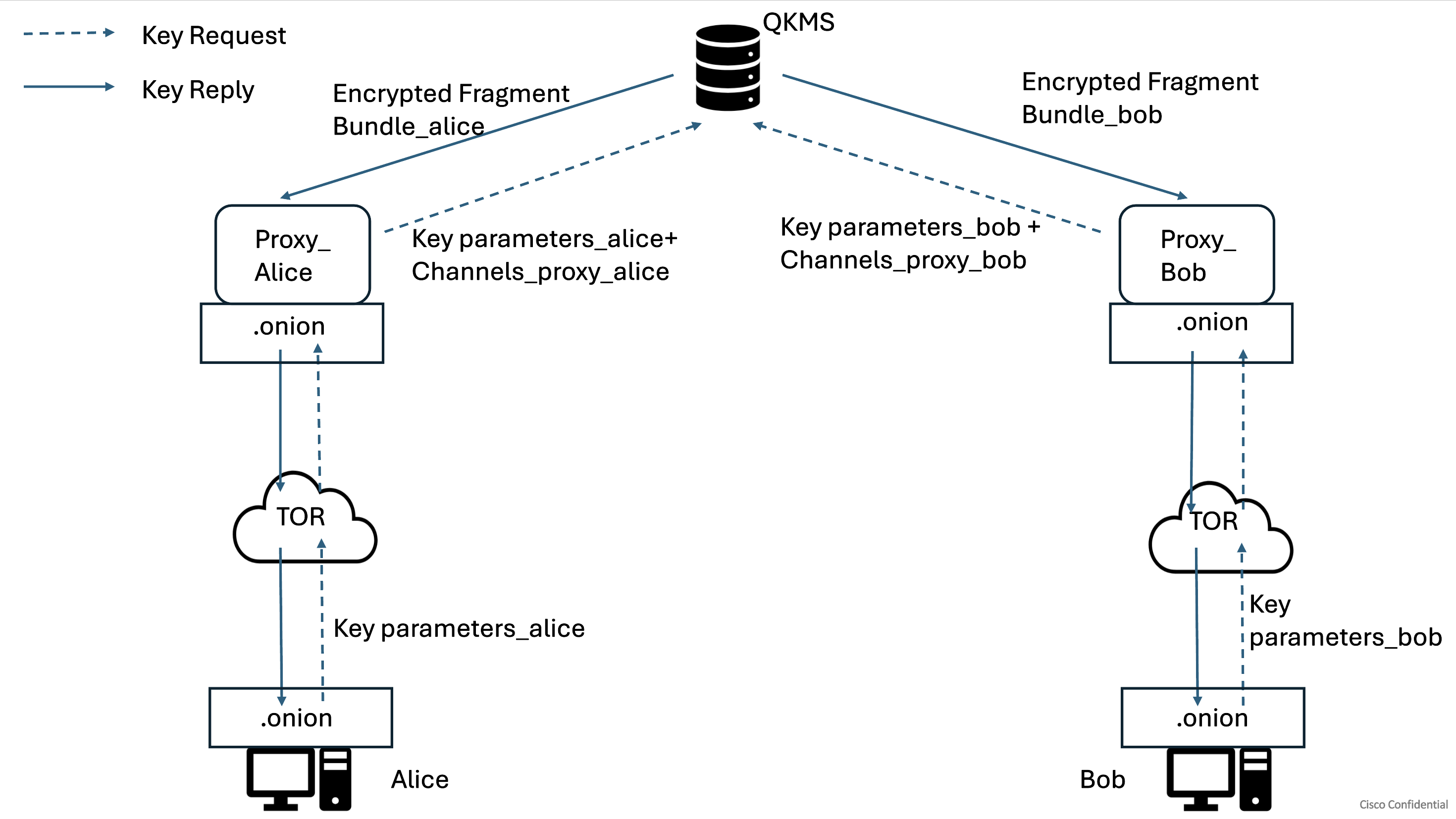}
  \caption{System architecture. Each client (Alice, Bob) hosts a Tor
           onion service and communicates with its proxy---also a Tor
           onion service---over the Tor network. The proxy receives
           encrypted fragment bundles from the QKMS over conventional
           channels and forwards each bundle to the client over an
           ephemeral Tor circuit obtained via the \texttt{NEWNYM} control
           signal. Dashed arrows denote key-request flow; solid arrows
           denote key-reply flow.}
  \label{fig:arch}
\end{figure*}

The system consists of three logical components: a Quantum Key Management
Server (QKMS) that generates session keys and partitions them into
encrypted fragments, communicating clients that request and reconstruct
keys, and onion-service \emph{proxies} that mediate between client and
QKMS over the Tor network. Each component is realized as a Flask
application exposing a small set of HTTP endpoints; cross-component
traffic between the client and the proxy traverses the Tor network as
onion-service traffic, while the proxy--QKMS traffic carries the
encrypted-fragment bundles over conventional channels. The overall
architecture is shown in Figure~\ref{fig:arch}.

\subsection{Quantum Key Management Server}
\label{sec:system:qkms}

The QKMS accepts a structured key-request message and replies with a
symmetric session key whose generation parameters are dictated by the
requester. Five parameters control the key-generation behavior; they are
included in the request body sent by each proxy on behalf of its client.

\textbf{tagname. } A unique identifier mutually agreed upon
        out-of-band by the two communicating clients. The QKMS proceeds
        only when the tagnames presented by the two endpoints match;
        this binds the issuance to a specific session and prevents
        unsolicited fragment delivery.
        
\textbf{key\_type. } The bit-length of the symmetric session
        key, sampled uniformly at random from $\{0,1\}^{\texttt{key\_type}}$.
        
\textbf{num\_of\_splits. } The number $n$ of fragments into
        which the key is partitioned. Reconstruction requires all $n$
        fragments; the scheme uses an $(n,n)$-style split rather than a
        $(t,n)$ threshold (see Section~\ref{sec:bg:ss}).
        
\textbf{shuffle. } A Boolean instructing the QKMS to randomly
        permute fragment order prior to dispatch. Combined with the
        per-fragment encryption, this prevents an observer that captures
        a single fragment from inferring its position in the original
        key.
        
\textbf{channels. } A list of logical channels exposed by the
        proxy, supplied by the proxy at request time. Each channel is a
        $\langle\text{host},\text{port}\rangle$ pair; the QKMS draws a
        channel for each fragment independently and uniformly. Some
        channels may carry zero fragments and others more than one.

After fragmentation and (optional) shuffling, each fragment is
encrypted under the recipient's public key (the client's classical RSA
key in the current prototype), and the encrypted fragments are dispatched
across the channel set as a small number of \emph{bundles}, each bundle
addressed to one channel. The QKMS exposes a single endpoint, POST
\texttt{/get-key}, that accepts the parameter set above together with the
requester's public key and triggers fragment generation and dispatch.

\subsection{Onion-Service Proxy}
\label{sec:system:proxy}

The proxy decouples the client from the QKMS and is the architectural
primitive by which the design supports clients that should not contact
the QKMS directly. The proxy is deployed as a Tor onion service: it is
addressed by a \texttt{.onion} hostname derived from its long-term
Ed25519 identity key (Section~\ref{sec:bg:tor}), and its IP address is
not exposed to the client. The proxy receives the client's payload---key
parameters and public key---over a Tor circuit, appends its own channel
list, and forwards the resulting payload to the QKMS via a POST request
to \texttt{/get-key}. Encrypted fragment bundles returned over the
channel set are aggregated at the proxy and forwarded to the client over
the Tor network. Because the fragments are encrypted under the client's
public key rather than the proxy's, the proxy never observes the session
key in plaintext.

\textbf{Per-bundle ephemeral circuits. }The design's anonymity property (Section~\ref{sec:security})
relies on each fragment bundle being forwarded over an \emph{independent}
Tor circuit. The proxy obtains a fresh circuit by sending a
\texttt{NEWNYM} signal to its local Tor process via the
\texttt{ControlPort}; subsequent stream attachments to the client's
\texttt{.onion} address are routed over a freshly sampled path through
the consensus (Section~\ref{sec:bg:tor}). The proxy issues one
\texttt{NEWNYM} signal per bundle, waits a short stabilization interval
to allow the new circuit to establish, and then issues the corresponding
POST request to the client's \texttt{/receive-key-fragment} endpoint.

We adopt the \texttt{NEWNYM} mechanism with two caveats from the Tor
documentation~\cite{stem_faq}. \emph{First}, \texttt{NEWNYM} causes Tor
to re-sample paths but does not guarantee node-disjoint circuits; we
accordingly treat the resulting circuits as approximately, not strictly,
independent. \emph{Second}, the Tor Project explicitly cautions against
aggressive circuit churn, since each \texttt{NEWNYM} imposes load on
volunteer relays. The protocol therefore aims for a modest $n$
(typically $n \in [2, 10]$ in our experiments) rather than an arbitrarily
large fragment count.

A representative \texttt{torrc} fragment for the proxy is shown in
Listing~\ref{lst:tor-config}. The \texttt{SocksPort} provides the local
entry point through which the proxy attaches its outbound streams; the
\texttt{ControlPort} (with cookie authentication) is the management
interface over which \texttt{NEWNYM} signals are issued; the
\texttt{HiddenServiceDir} and \texttt{HiddenServicePort} directives
configure the proxy itself as an onion service reachable on the
corresponding \texttt{.onion} address.

\begin{listing}[t]
\caption{Representative proxy \texttt{torrc} excerpt. Specific port
         numbers, country selectors, and paths vary per deployment.}
\label{lst:tor-config}
\begin{minted}[fontsize=\small, breaklines, linenos, xleftmargin=2em,
               numbersep=5pt]{bash}
SocksPort 8005
ControlPort 8006
CookieAuthentication 1

EntryNodes {us},{ca},{de}
ExitPolicy reject *:*

HiddenServiceDir   /var/lib/tor/proxy/hs
HiddenServicePort  5000 127.0.0.1:5000

DataDirectory      /var/lib/tor/proxy/data
Log                notice file /var/log/tor/proxy.log
\end{minted}
\end{listing}

\subsection{Client}
\label{sec:system:client}

The client also runs as a Tor onion service so that the proxy never
learns the client's IP address; addressing between proxy and client is
done by \texttt{.onion} hostname in both directions. The client
initiates a key request by issuing a POST to the proxy's
\texttt{/get-key} endpoint via the local Tor SOCKS proxy
(Listing~\ref{lst:client-payload}), and listens for fragment deliveries
on its own \texttt{/receive-key-fragment} endpoint. On receipt, the
client (i) decrypts each fragment with its private key,
(ii) checks that the expected number of fragments has arrived,
(iii) undoes the shuffle order using the per-fragment index, and
(iv) concatenates the fragments to reconstruct the session key. The
reconstructed key is then used as a master secret for AES-GCM
authenticated encryption of application traffic.

\begin{listing}[t]
\caption{Client and proxy payloads. The client payload contains the
         parameter set and the client's public key; the proxy augments
         this with its own channel list before forwarding to the QKMS.}
\label{lst:client-payload}
\begin{minted}[fontsize=\small, breaklines, linenos, xleftmargin=2em,
               numbersep=5pt]{js}
// Client -> Proxy (POST /get-key)
{
  "tagname":       "session-2026-05-12-001",
  "key_type":      768,
  "num_of_splits": 10,
  "shuffle":       true,
  "public_key":    "<client RSA public key>"
}

// Proxy -> QKMS (POST /get-key)
{
  "tagname":       "session-2026-05-12-001",
  "key_type":      768,
  "num_of_splits": 10,
  "shuffle":       true,
  "public_key":    "<client RSA public key>",
  "channels": [
    "http://10.0.0.5:4000/",
    "http://10.0.0.5:4001/"
  ]
}
\end{minted}
\end{listing}

\subsection{Protocol Flow}
\label{sec:system:flow}

The end-to-end flow, illustrated in Figure~\ref{fig:arch}, is as follows.
\begin{enumerate}
  \item Each of two clients $\{A,B\}$ agrees out-of-band on a common
        \texttt{tagname} and learns the \texttt{.onion} address of its
        assigned proxy.
  \item Each client issues POST \texttt{/get-key} to its proxy over Tor
        with the parameter set of Listing~\ref{lst:client-payload}.
  \item Each proxy appends its channel list and forwards the request to
        the QKMS over a conventional channel.
  \item Upon receiving matching tagnames from both proxies, the QKMS
        generates the session key, partitions it into $n$ fragments,
        encrypts each under the corresponding client's public key,
        optionally shuffles them, and dispatches one or more bundles to
        each proxy's channel set.
  \item Each proxy aggregates the received bundles and, for each bundle,
        issues a \texttt{NEWNYM} signal, waits for circuit stabilization,
        and forwards the bundle to the client's \texttt{.onion} address
        via a POST to \texttt{/receive-key-fragment}.
  \item Each client decrypts the received fragments, validates the
        expected count, undoes the shuffle, and concatenates to recover
        the session key.
\end{enumerate}

A schematic of the reconstructed key fragments produced by the QKMS is
shown in Listing~\ref{lst:key-fragments}; for brevity, only the
first three fragments of a typical $n=10$ session are reproduced.

\begin{listing}[t]
\caption{Representative fragment output (first three of $n=10$). The
         shuffle order is preserved as a part-index so that the client
         can reassemble the key in the correct order after decryption.}
\label{lst:key-fragments}
\begin{minted}[fontsize=\small, breaklines, linenos, xleftmargin=2em,
               numbersep=5pt]{text}
Part 7 of 10: 1111000101010100010010100...
Part 1 of 10: 1010100100001111001001010...
Part 2 of 10: 101010000110101000000001...
...
\end{minted}
\end{listing}

The above flow makes explicit the two structural choices that distinguish
this design from a single-circuit Tor tunnel: (i) the key is partitioned
across $n$ bundles, each carried on its own ephemeral circuit, and
(ii) both proxy and client are addressed as onion services, so each
fragment traverses the onion-service rendezvous protocol---a six-hop
path with three hops from each side meeting at a shared rendezvous
point~\cite{torproject_hs}, rather than the three-hop circuit used to
reach a clearnet destination. The first choice provides the geometric
linkability bound of Proposition~\ref{prop:decay}; the second provides
the bidirectional anonymity (neither endpoint learns the other's IP)
that is the precondition for that bound to be the right object to
bound.
\section{Security Analysis}
\label{sec:security}

Having described the system in Section~\ref{sec:system}, we now state
the security property the design provides and argue why it is achieved
under the adversary of Section~\ref{sec:threat:adversary}. The argument
rests on already-established bounds from the onion-routing
literature~\cite{dingledine2004,murdoch2007} rather than on any new
cryptographic assumption.

\subsection{Security Property}
\label{sec:security:property}

We aim for the following property.

\begin{definition}[Anonymity-Preserving Reconstruction]
\label{def:apr}
A multi-fragment key-distribution protocol provides
\emph{anonymity-preserving reconstruction} against an adversary
$\mathcal{A}$ if, after observing the network traffic produced by a
session, $\mathcal{A}$ cannot link the $n$ ciphertext fragments
belonging to that session to a common pair of endpoints with
probability non-negligibly greater than uniform guessing.
\end{definition}

The property is independent of the secrecy of the per-fragment
payload: an HNDL adversary that captures and later decrypts a single
fragment recovers \emph{that fragment's} plaintext. What our scheme
provides is that, in the absence of successful end-to-end correlation
across \emph{every} Tor circuit involved in the session, an HNDL
adversary cannot \emph{group} the decrypted fragments into a single
key reconstruction event with attribution to a specific client.

\subsection{Per-Circuit Correlation Bound}
\label{sec:security:single}

We adopt the standard Tor analysis~\cite{dingledine2004,murdoch2007}.
Suppose $\mathcal{A}$ surveils $\mathcal{S} \subseteq \mathcal{R}$ with
$|\mathcal{S}|/|\mathcal{R}| = f$. For a single onion-service circuit
consisting of two halves, each a three-hop circuit constructed by an
independent endpoint to a common rendezvous point, the probability
that $\mathcal{A}$ observes both the client-side guard relay and the
service-side guard relay is, under independent sampling of relays from
$\mathcal{R}$,
\begin{equation}
  \label{eq:single-circuit}
  \Pr[\textsf{corr}_1] \;\approx\; f^2
\end{equation}
for the well-studied entry/exit correlation
attack~\cite{dingledine2004,murdoch2007,nasr2018deepcorr}. The $f^2$
bound is the standard simplification under which the adversary
observes both ends of \emph{one} 3-hop half; observing both halves
simultaneously requires additional assumptions and is bounded by $f^2$
in the same regime under the same model. We use $f^2$ as the
conservative per-circuit success probability and treat any tighter
analysis as future work.

\subsection{Multi-Circuit Composition}
\label{sec:security:composition}

The protocol creates $n$ ephemeral Tor circuits, one per fragment
bundle, by issuing a \texttt{NEWNYM} control signal between bundles.
Two empirical caveats apply. \emph{First}, \texttt{NEWNYM} causes Tor
to re-sample paths from the current consensus, but it does not
guarantee node-disjoint circuits; on a relay set of size $P$ with the
guard selection algorithm, two consecutive circuits will reuse some
relays with non-zero probability~\cite{stem_faq}. \emph{Second}, the
per-session guard pinning policy in modern Tor reuses a small set of
long-lived guards for an onion service, which biases the entry side
across circuits; this is an explicit design choice for vanilla onion
services~\cite{torproject_hs}. Subject to these caveats, the design
treats the $n$ circuits as \emph{approximately independent} draws from
the population $\mathcal{R}$ of relays. Under this approximation, the
probability that $\mathcal{A}$ correlates the entry and exit relays of
\emph{every} one of the $n$ circuits is
\begin{equation}
  \label{eq:multi-circuit}
  \Pr[\textsf{corr}_{1..n}] \;\approx\; \bigl(\Pr[\textsf{corr}_1]\bigr)^{n}
  \;\approx\; f^{2n}.
\end{equation}

\begin{proposition}[Linkability decays geometrically in $n$]
\label{prop:decay}
Under the approximate independence of fresh Tor circuits (with the
caveats noted above) and the standard per-circuit correlation bound
$\Pr[\textsf{corr}_1] \approx f^2$, the probability that the adversary
$\mathcal{A}$ correlates every one of the $n$ ephemeral circuits
carrying the $n$ fragments of a single session is bounded by $f^{2n}$.
\end{proposition}

\textbf{Interpretation. }For concreteness, take $f = 0.05$---a generous upper bound on the
fraction of Tor relays that any one adversary is plausibly believed to
control today---and $n = 10$. Then $f^{2n} = 0.05^{20} \approx
10^{-26}$, which is comfortably below any operational threshold. Even
much larger $f$ values yield very small composed correlation
probabilities for modest $n$. The point of this calculation is not to
claim a numeric guarantee but to make explicit that the design's
anonymity property inherits geometrically from the per-circuit Tor
bound and the choice of $n$.

\subsection{Scope of the Bound}
\label{sec:security:scope}

We emphasize three things the bound in Equation~\ref{eq:multi-circuit} does
\emph{not} claim.
\begin{enumerate}
  \item It does not provide post-quantum confidentiality of any one
        fragment's plaintext payload: per-fragment encryption is
        classical, so an HNDL adversary that captures a single
        fragment and later acquires a CRQC recovers its plaintext.
  \item It is not tight under non-independence: shared guards, shared
        middle relays through Tor's guard-selection algorithm, or
        time-correlated path selections all reduce the effective
        independence of the $n$ circuits, and the true correlation
        probability is generally larger than $f^{2n}$. See
        Section~\ref{sec:discussion} for further discussion.
  \item It is not a defense against confirmation attacks
        (Section~\ref{sec:threat:nongoals}): an adversary with strong
        out-of-band evidence that endpoint $C$ communicates with QKMS
        at a given time may verify that hypothesis with much less
        effort than the bound suggests.
\end{enumerate}
The geometric-decay bound is a \emph{linkability} property suitable
for the HNDL setting, where the adversary's primary task is to
attribute future decryptions to specific endpoint pairs.
\section{Implementation and Evaluation}
\label{sec:eval}

We implement the design as a Python/Flask prototype with Tor integration
via the \texttt{stem} controller library~\cite{stem} and measure
end-to-end key-establishment latency on a small AWS EC2 testbed. The
goal of the evaluation is not to claim a production-ready latency profile
but to (i) demonstrate that the protocol completes correctly end-to-end
over the public Tor network and (ii) decompose the observed latency
into its cryptographic and network-related components so that the
contribution of the Tor-routing layer is explicit.

\subsection{Testbed}
\label{sec:eval:testbed}

The testbed consists of five entities. Two communicating clients run on
separate physical laptops (Ubuntu 24.04, Intel Core i7-12800H, 32\,GB
RAM). The QKMS and two intermediary proxy nodes (one per client--QKMS
pair) are deployed as independent Amazon EC2 instances running
Ubuntu 24.04. Each entity runs the Flask-based implementation of
Section~\ref{sec:system}. Tor is configured on every entity with the
\texttt{SocksPort} acting as the local entry to the Tor network and a
\texttt{ControlPort} (cookie authentication) over which \texttt{NEWNYM}
signals are issued. Both the proxies and the clients are deployed as
onion services as described in Section~\ref{sec:system:proxy}.

\subsection{Methodology}
\label{sec:eval:method}

We measure the end-to-end latency of a complete key establishment from
the moment a client issues its POST \texttt{/get-key} request to the
moment it has reconstructed the session key from all $n$ decrypted
fragments. Two channels (logical port-bound endpoints on the proxy) are
instantiated per proxy in the current prototype; the channels carry
random subsets of the $n$ fragments per session, and each bundle is
forwarded to the client over its own fresh Tor circuit
(Section~\ref{sec:system:proxy}). To respect the Tor Project's published
guidance on circuit churn~\cite{stem_faq}, we deliberately constrain the
volume and frequency of test runs and report the resulting measurements
with the corresponding caveat
(Section~\ref{sec:eval:threats}).

A representative decryption trace at the client is reproduced in
Listing~\ref{lst:client-dec}, showing the per-fragment decryption time
and the part-index used for reordering.

\begin{listing}[t]
\caption{Representative client-side decryption trace (excerpted; ciphertext
         truncated for legibility). Each fragment is decrypted with the
         client's RSA private key in $\sim 0.2$\,s, and the recovered
         part-index drives reassembly.}
\label{lst:client-dec}
\begin{minted}[fontsize=\small, breaklines, linenos, xleftmargin=2em,
               numbersep=5pt]{text}
[CLIENT] Received share (idx=2): RXCtLzxrQyC95hLoZL7ASHN...
Decrypted: part 2 of 10, 1001100100010100001110100...
Decryption time: 0.199 s

[CLIENT] Received share (idx=3): wAYE9Vp+M3xqQK9LzAaQ0r...
Decrypted: part 3 of 10, 1100100010011101110111000...
Decryption time: 0.185 s

[CLIENT] All 10 fragments received; reassembling.
\end{minted}
\end{listing}

\subsection{Results}
\label{sec:eval:results}

Over the test runs we performed, the complete end-to-end interaction
took 13--20\,s on average, with the full observed range spanning
7--50\,s. The lower end of the range corresponds to favorable Tor
conditions (low circuit-build times, no relay churn during the request);
the upper end corresponds to runs in which one or more
\texttt{NEWNYM}-induced circuit constructions encountered congestion or
relay re-selection. The variance is consistent with the latency
characteristics of onion-service rendezvous reported in the Tor
performance literature~\cite{wilms2008,darkhorse2023}.

\textbf{Decomposition. }We attribute approximately 88\% of the observed end-to-end latency to
Tor-related delay, with the remaining $\sim 12\%$ split between
per-fragment classical-public-key encryption at the QKMS, per-fragment
classical-public-key decryption at the client, key sampling, fragment
shuffling, and Flask request handling. The dominant Tor-related
contribution is the per-bundle circuit construction triggered by
\texttt{NEWNYM}: each bundle pays the cost of stabilizing a fresh
client-to-rendezvous-to-service path before the bundle's POST request
can be issued. The cryptographic work itself is negligible at the scales
considered here; this matches the well-known result that
public-key-encryption cost is small relative to interactive Tor latency
on commodity hardware~\cite{berger2025tor}.

\textbf{Behavior in the number of fragments. }The end-to-end latency scales approximately linearly in $n$, since each
fragment incurs its own circuit-build delay. The protocol is therefore
most useful in regimes where $n$ is small and the per-session anonymity
budget---roughly $f^{2n}$ in the bound of
Proposition~\ref{prop:decay}---is already well below the operational
threshold. For the configurations we tested ($n \in \{2, 10\}$), the
absolute end-to-end latency is dominated by Tor and the protocol overhead
beyond the cryptographic work is negligible.

\subsection{Threats to Validity}
\label{sec:eval:threats}

Three caveats apply to the empirical results above. \emph{First}, we
report mean values from a deliberately small number of test runs in
order to respect the Tor Project's published guidance on the load
imposed by aggressive circuit churn~\cite{stem_faq}; a percentile
breakdown---particularly the 95th and 99th percentiles of the per-bundle
circuit-build time---would provide a more complete operational picture
and is left to future work. \emph{Second}, the current prototype
exercises a two-channel configuration, and a larger $n$ might surface
non-linearities in the per-bundle circuit-build cost that we do not
observe here. \emph{Third}, all measurements are taken on the public Tor
network during specific time windows; load conditions on Tor vary
substantially over time and our measurements are not corrected for
diurnal load patterns. None of these caveats affects the qualitative
conclusion---that Tor dominates the latency---but each constrains the
precision of the reported numbers.

\section{Discussion}
\label{sec:discussion}

The empirical results of Section~\ref{sec:eval} make a single
qualitative point: the price of the anonymity property formalized in
Proposition~\ref{prop:decay} is paid almost entirely in Tor-related
latency, and the cryptographic and bookkeeping overheads of the
multi-fragment design are negligible by comparison. The implication is
that the design's deployability is gated by the user's tolerance for
13--20\,s end-to-end establishment time (with non-negligible tails to
50\,s), not by its cryptographic or protocol mechanics.

\textbf{Privacy-versus-responsiveness trade-off. }The decision to construct a fresh Tor circuit for every fragment bundle
is the dominant source of latency in the prototype. The trade-off it
encodes is the classical one in anonymity systems: per-bundle circuit
freshness reduces the long-term observability of any single circuit and
therefore reduces the probability that a network adversary can perform
the kind of multi-circuit timing correlation that would link the
fragments back to a common pair of endpoints. The cost is the
circuit-build latency itself, which is irreducible on top of the
six-hop onion-service rendezvous protocol. For interactive workflows,
20\,s is at or beyond the threshold of usability, and for
latency-sensitive applications the design is plainly inappropriate. For
the use cases the threat model targets---HNDL-resistant session-key
establishment for data with multi-year confidentiality
lifetimes---a one-time per-session 13--20\,s overhead is a small price
to pay if the alternative is decryption of archived ciphertext
years hence.

\textbf{Approximate independence of fresh circuits. }The geometric-decay bound of Proposition~\ref{prop:decay} relies on
treating the $n$ ephemeral circuits as approximately independent draws
from the Tor relay population. This approximation breaks in two ways
that matter operationally.
\begin{itemize}
  \item \emph{Guard reuse.} Modern Tor pins a small set of long-lived
        guard relays per service; \texttt{NEWNYM} re-samples middle and
        exit relays but generally does not change the guard. The
        entry-side of the $n$ circuits is therefore correlated, not
        independent, and the true correlation probability is larger than
        $f^{2n}$ in the regime where the guard is itself in
        $\mathcal{S}$.
  \item \emph{Bandwidth-weighted sampling.} Path selection in Tor is
        weighted by relay-advertised bandwidth, so high-bandwidth relays
        appear in many circuits. An adversary that controls a
        high-bandwidth relay therefore has higher per-circuit observation
        probability than a uniform-sampling analysis would suggest.
\end{itemize}
Both effects can be partially mitigated---one can deliberately rotate
guards (with corresponding security implications of its own), or use
a relay-selection policy that is robust to bandwidth-weighting
attacks---but the analysis as presented should be understood as an upper
bound rather than as a tight worst-case guarantee. We regard tightening
the bound under realistic Tor guard and bandwidth-weighting behavior as
the most important piece of follow-on theoretical work for this line.

\textbf{Confirmation versus correlation. }The bound in Proposition~\ref{prop:decay} addresses \emph{correlation}:
the question of whether an adversary observing many circuits can pick
out the $n$ that belong to a single session. It does not address
\emph{confirmation}: the question of whether an adversary that already
hypothesizes a specific pair of endpoints can verify that
hypothesis~\cite{torproject_hs}. An adversary with strong out-of-band
evidence---e.g., a court order requiring an ISP to log a specific
client's traffic---can confirm communication with much less effort than
the geometric bound implies. We make no claim of resistance to
confirmation attacks; the design's anonymity guarantee is appropriate
to the HNDL setting, in which the adversary's primary task is
\emph{attribution} of a future decryption to a specific endpoint pair,
not verification of a pre-existing hypothesis.

\textbf{Ethical considerations. }We deliberately constrained the volume and frequency of our test runs
to avoid placing undue load on the public Tor network, which is a
volunteer-operated and rate-limited shared resource. The Tor Project's
\texttt{stem} documentation explicitly cautions against aggressive
circuit churn~\cite{stem_faq}, and the per-bundle \texttt{NEWNYM} of
the protocol is precisely the workload pattern that incurs the most
load. A production deployment of the protocol should either run its
own Tor-instance with bandwidth contribution proportional to its
consumption, or move to a private deployment of onion routing rather
than the public Tor network. Until that step is taken, the protocol
should be regarded as a research prototype.

\textbf{Limitations. }Beyond the validity threats noted in Section~\ref{sec:eval:threats}, we
note three structural limitations of the prototype.
\begin{enumerate}
  \item The per-fragment encryption uses RSA, which is classically
        secure today but quantum-broken in the HNDL window the design
        targets. An adversary that captures every fragment and later
        acquires a CRQC recovers every fragment's plaintext; the
        protocol resists \emph{linkability}, not per-fragment
        confidentiality, in this regime. A hybrid envelope construction
        (asymmetric key-wrap of a per-fragment symmetric key, with the
        key-wrap performed using a hybrid KEM~\cite{westerbaan2023,
        torspec269}) would close this gap, and is the most important
        deployment hardening for any non-research use of the design.
  \item The current prototype dispatches at most a handful of fragments
        and exercises a two-channel configuration; the linear scaling in
        $n$ of end-to-end latency means that larger $n$, while
        cryptographically attractive (Proposition~\ref{prop:decay}),
        is operationally costly.
  \item Bootstrap of the proxy's \texttt{.onion} address to the client is
        out of scope for this paper; we assume that the client has
        obtained the proxy's \texttt{.onion} address through an
        out-of-band, quantum-safe mechanism (e.g., an NFC kiosk
        bootstrap, a pre-provisioned token, or an in-person handoff).
        Designing a quantum-safe bootstrap that does not itself rely on
        classical asymmetric cryptography is an interesting design
        problem that we do not address here.
\end{enumerate}

\textbf{Composition with hybrid KEMs. }Nothing in the protocol precludes also performing a hybrid PQ KEM on the
wire-line tunnel between the proxy and the QKMS, or between the proxy
and the client. Such a composition would give a two-property defense:
the hybrid KEM provides per-fragment confidentiality even under an HNDL
adversary that captures every fragment, while the multi-circuit Tor
layer provides the linkability property of
Proposition~\ref{prop:decay}. The two layers are structurally
independent, and we regard their joint deployment as the most natural
hardening of the design for production use.

\section{Conclusions}
\label{sec:conclusion}

We have presented a session-key establishment scheme that distributes a
symmetric key as $n$ independently encrypted fragments across distinct
ephemeral Tor circuits between an onion-service proxy and an
onion-service client, with a security argument grounded in the standard
end-to-end correlation bound for onion
routing: under the approximate
independence of fresh Tor circuits, an adversary surveilling a fraction
$f$ of relays correlates every one of the $n$ circuits of a session with
probability $f^{2n}$ (Proposition~\ref{prop:decay}). Our AWS EC2
prototype completes a key establishment in 13--20\,s end-to-end, with
roughly 88\% of the latency attributable to Tor; the cryptographic and
fragmentation work is negligible by comparison, so the design is
practical at the protocol level and bounded operationally only by the
user's tolerance for per-bundle circuit-build latency.

\bibliographystyle{IEEEtran}
\bibliography{main}

\end{document}